\documentclass[final]{iucrjournals}

\usepackage{listings}
\usepackage{xcolor}

\lstdefinestyle{clistyle}{
  basicstyle=\ttfamily\small,
  backgroundcolor=\color{black!4},
  frame=single,
  framerule=0pt,
  xleftmargin=6pt,
  xrightmargin=6pt,
  aboveskip=6pt,
  belowskip=6pt,
  breaklines=true,
  columns=fullflexible,
  keepspaces=true,
  showstringspaces=false,
}
\title{EQSANS-CLI: A natural-language, agent-ready command-line tool for small-angle neutron scattering data reduction at EQ-SANS}

\author[a]{Changwoo Do\IUCrCemaillink{doc1@ornl.gov}\IUCrOrcidlink{0000-0001-8358-8417}}

\affil[a]{Neutron Scattering Division, Oak Ridge National Laboratory, 1 Bethel Valley Road, Oak Ridge, TN 37831, USA}

\begin{document}
\nolinenumbers

\begin{center}
\textbf{NOTICE OF COPYRIGHT}\\
This manuscript has been authored by UT-Battelle, LLC under Contract DE-AC05-00OR22725 with the U.S. Department of Energy (DOE). The U.S. government retains and the publisher, by accepting the article for publication, acknowledges that the U.S. government retains a nonexclusive, paid-up, irrevocable, worldwide license to publish or reproduce the published form of this manuscript, or allow others to do so, for U.S. government purposes. DOE will provide public access to these results of federally sponsored research in accordance with the DOE Public Access Plan (\url{http://energy.gov/downloads/doe-public-access-plan}).
\end{center}
\clearpage

\maketitle

\begin{synopsis}
EQSANS-CLI is a command-line tool for small-angle neutron scattering data reduction that accepts slash commands, natural language, and a headless JSON protocol through a shared command-handler layer, making the full reduction pipeline addressable by users and by external AI agents.
\end{synopsis}

\begin{abstract}
Small-angle neutron scattering (SANS) data reduction at user facilities follows a largely repetitive workflow. Runs are first classified in the catalog and matched to transmission, background, and empty-beam references within the same instrument configuration. The data are then reduced, placed on an absolute scale using a standard, and stitched across configurations. Although these steps are individually well understood, they remain weakly connected, producing a coordination burden that scales with the number of runs and configurations. This paper describes EQSANS-CLI, a command-line tool that organizes this workflow into a coherent, scriptable, and agent-addressable system. The design rests on four principles: a shared command-handler layer that all input paths converge on; a persistent \emph{working table} that holds every reduction decision as editable rows; two input surfaces (an interactive terminal and a headless JSON mode) that compile to the same handler entry point; and a status-driven re-reduction model that treats parameter changes as first-class events. An \texttt{/autopilot} command chains the full pipeline from the IPTS number to stitched $I(Q)$ curves in one invocation. A Slack bot demonstrates that the headless interface, together with a single skill document loaded into an external agent's system prompt, is sufficient to drive complete reductions by natural language from a mobile device. The architecture is intentionally minimal on the agent side: the CLI is the authoritative executor, and the agent's only job is to translate human intent into commands on a stable contract.
\end{abstract}


\section{Introduction}

Small-angle neutron scattering (SANS) is an indispensable probe of structure on length scales from nanometres to sub-micrometres in soft matter, biomaterials, and energy materials \cite{heller2018sans}. The Extended Q-range Small-Angle Neutron Scattering (EQ-SANS) instrument at the Spallation Neutron Source (SNS) \cite{zhao2010eqsans} has operated as a workhorse for the soft-matter community since 2009, and its workflow is representative of modern SANS beamlines: a researcher proposes an experiment, selects one or more instrument configurations to cover the desired $Q$-range, writes an acquisition script for the EPICS-based control system \cite{dalesio1994epics,yao2021unified}, collects scattering and transmission data for each sample in each configuration, reduces the raw event data to $I(Q)$, stitches configurations together, and finally analyzes the result. Each of these stages has seen substantial automation in isolation. Instrument control has evolved from instrument-specific control systems to facility-wide platforms such as EPICS and Control System Studio \cite{yao2021unified}. Python scripting wrappers simplify common acquisition patterns. 

More recently, advances in artificial intelligence, particularly large language models (LLMs), have begun to further extend these automation efforts by bridging human intent and instrument control. An assisting chatbot, ESAC, translates natural-language experimental plans into complete acquisition scripts for EQ-SANS \cite{do2025esac,do2025esacv1}. On the analysis side, mature libraries such as SasView \cite{sasview} have been augmented by LLM-based agents that provide a natural-language interface to model selection, fitting, and interpretation \cite{ding2026sasagent}. 

Data reduction, however, has remained the stage most resistant to simplification. A typical EQ-SANS experiment produces 50--200 runs across two or three configurations. Each scattering run must be classified (scattering, transmission, background, empty-beam), matched to its corresponding transmission and background references within the same configuration, reduced with the correct sample thickness and absolute-scale factor, and stitched across configurations over a carefully chosen overlap region. Upwards of fifty configurable parameters are involved in a single reduction. The process is mechanical but not trivial. In practice, it has been carried out by editing standalone Python scripts that encode the full reduction workflow. While effective, this approach requires users to manage many parameters explicitly and to maintain consistency across runs and configurations. As the number of runs and configurations increases, the coordination effort grows quickly, making the workflow difficult to scale and prone to user error.

The response pursued in this paper is not a new reduction pipeline---EQ-SANS already has a mature reduction engine in \texttt{drtsans}---but a new interface to the existing pipeline that can be used in three ways: from a terminal, through plain English, and by an external AI agent over a network. The third case is motivated by a simple observation, increasingly supported across user facilities \cite{beaucage2023afl,sutherland2024autosas,xiao2024neudiff,li2025agentic,moreno2026hep,prince2023opportunities}. Large language models can handle much of the translation between user intent and instrument-specific commands when they are given a stable programmatic interface. The main contribution of this work is the design of such an interface.

This paper describes EQSANS-CLI, a command-line tool for EQ-SANS data reduction. It provides a single command-handler layer that is accessed through two input modes: an interactive terminal UI and a headless JSON mode designed for external agents. An \texttt{/autopilot} command chains the full reduction pipeline—catalog loading, run matching, calibration, reduction, and stitching—into a single invocation. 
A Slack bot is used to demonstrate how this interface can be operated by an external AI agent. The bot acts as a minimal client, with no EQ-SANS-specific logic beyond a skill document loaded into its system prompt, and can drive the full reduction workflow through natural-language interaction from a mobile device. More importantly, this example shows that an agent-ready interface allows users to connect their own AI agents to the system. Such agents can potentially automate not only data reduction, but also downstream tasks such as analysis and discussion. This suggests a broader direction in which scientific software is designed from the outset to support AI agents.
The source code is available at \url{https://github.com/cw-do/eqsanscli} \cite{eqsanscli}.

The remainder of the paper is organized as follows. Section~2 presents the design principles of EQSANS-CLI and explains why the natural-language and headless interfaces share a common handler layer. Section~3 describes the architecture, with emphasis on the working table as the central state, the command-handler layer, the natural-language interface, configuration identity, and the status-driven re-reduction model. Section~4 describes the user workflow, including the \texttt{/autopilot} pipeline and the smart stitching algorithm. Section~5 presents the headless mode and the Slack agent demonstration. Section~6 discusses limitations and future directions.

\section{Design}

EQSANS-CLI is designed around four principles. These principles determine the architecture and explain most implementation choices. They are stated here as simple claims about how the system behaves. Their realization is described in Section~3.

\emph{Principle 1: A single command-handler layer executes all operations.}
Every operation that changes the system state is implemented exactly once as a slash command. Both input modes—the interactive terminal and the headless JSON interface—dispatch commands through the same handler layer. 
In the terminal, users can type commands directly or enter natural-language input, which is translated into the same commands before execution. External agents interact through the headless interface by sending commands over a structured protocol. In all cases, the same handlers are used, and there are no parallel code paths that execute side effects.
This design ensures that all state-changing operations follow a single, consistent execution path. It also keeps the system auditable, since the point of execution is always the same regardless of how the command was generated.

\emph{Principle 2: The working table is the authoritative record of the reduction.} Every reduction decision is stored as a row in a single in-memory table. This includes run assignments, configuration, parameters, output location, and status. The table persists across commands and across sessions. 
The working table answers a simple question: what will happen when reduction is executed? Because it is explicit, the state can be inspected at any time using \texttt{/show table}. It can also be modified using the same commands that created it. This makes targeted edits straightforward and avoids the need to reconstruct state from scratch.

\emph{Principle 3: Status-driven re-reduction.} A common source of error in SANS reduction is re-running outdated results after a parameter change. EQSANS-CLI tracks status at the row level. A row moves from \texttt{ready} to \texttt{done} after reduction. It moves to \texttt{modified} when any relevant parameter changes. 
By default, the next reduction re-runs only \texttt{modified} rows and skips those marked \texttt{done}. This makes iterative work efficient. It also makes \texttt{/autopilot} resumable from a partial state. This behavior is important for both interactive use and agent-driven workflows, where sessions may be interrupted and resumed.

The rest of the paper describes how these principles are realized in the system.

\section{Architecture}

\subsection{System structure}
\begin{figure}[!htbp]
\centering
\includegraphics[width=0.9\textwidth]{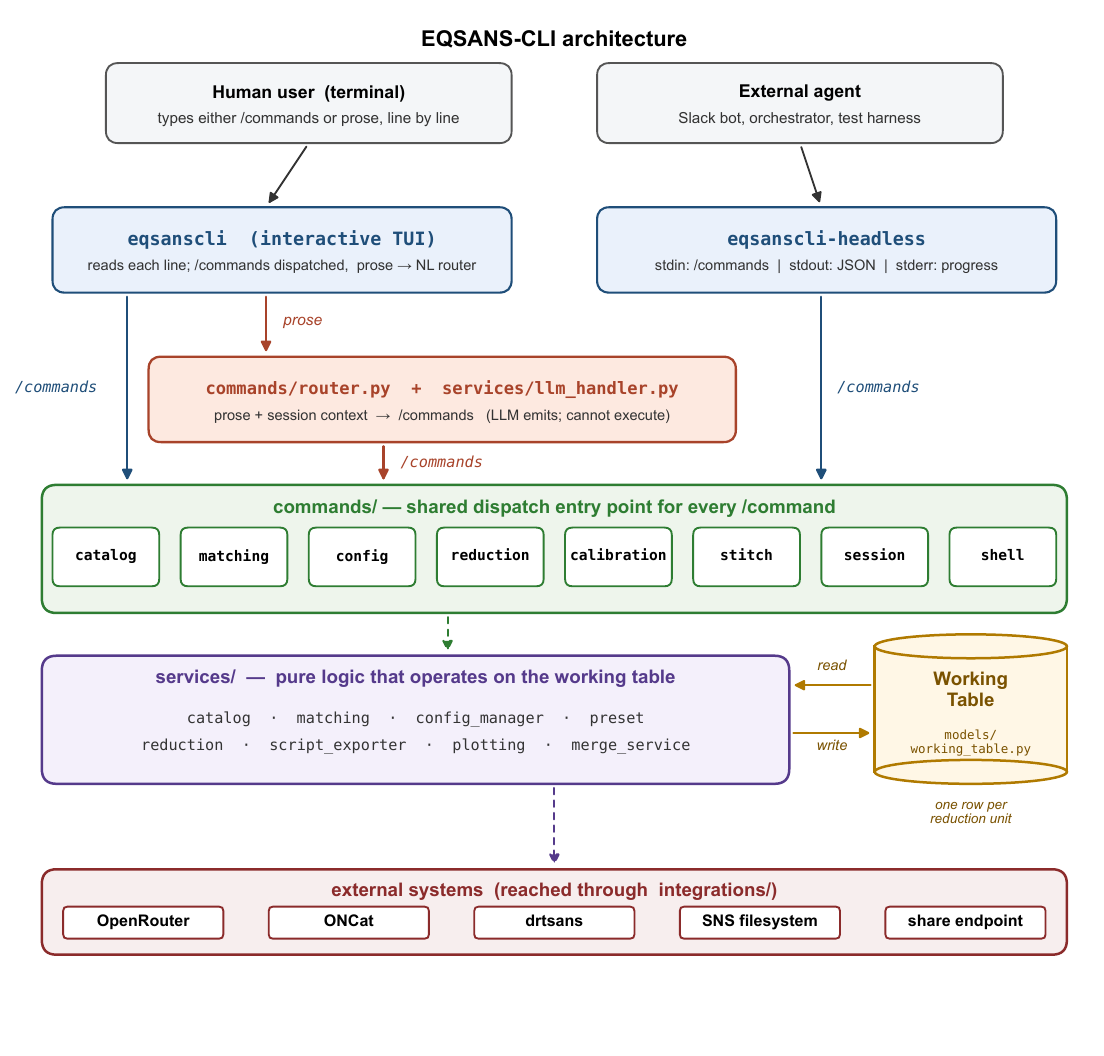}
\caption{Architecture of EQSANS-CLI. Two client types reach the tool through two binaries: a human uses the interactive TUI, and an external agent uses the headless JSON engine. Both paths dispatch commands through the same handler layer. The TUI can also pass free-form prose to a natural-language route that compiles it into slash commands. Handlers call services that operate on the working table and reach external systems through integration modules.}
\label{fig:architecture}
\end{figure}

Figure~\ref{fig:architecture} summarizes the layered architecture. Two types of client reach the tool: a human user at a terminal, and an external AI agent connecting over SSH. They feed two binaries respectively: the interactive TUI (\texttt{eqsanscli}) and the headless JSON engine (\texttt{eqsanscli-headless}). The TUI reads each line of input and routes it in one of two ways. Lines that start with a forward slash are dispatched directly to the shared command-handler layer (\texttt{commands/}), the same layer used by the headless engine. Lines that do not start with a slash are treated as natural-language input and passed to a dedicated natural-language route (\texttt{commands/router.py} together with \texttt{services/llm\_handler.py}), which assembles a prompt from curated domain knowledge and the current session context, asks an LLM to return a sequence of slash commands, and then dispatches those commands through the same handler layer as everything else. The user can switch between typed commands and prose on a line-by-line basis. There is no separate “natural-language mode.” The LLM only translates prose into slash commands. It does not call reduction functions or modify the working table directly.

The command handlers in \texttt{commands/} are grouped by concern (catalog, matching, config, reduction, calibration, stitch, session, shell). They call into pure-logic modules in \texttt{services/} (catalog, matching, config manager, preset, reduction, script exporter, plotting, merge service). 
The services operate on a shared in-memory \emph{working table} held in \texttt{models/working\_table.py}. They also reach external systems through wrapper modules in \texttt{integrations/}. 
The external systems include ONCat \cite{oncat} for run metadata. They also include the SNS filesystem (\texttt{/SNS/EQSANS/...} on the analysis cluster) for raw NeXus event data, reduced $I(Q)$ files, and merged outputs. The reduction engine is \texttt{drtsans}, which is invoked as a subprocess. A public share endpoint \cite{herenow} provides time-limited URLs for plots and reduced files. OpenRouter is used as the LLM provider for the natural-language route. 
Of these, only OpenRouter is specific to the natural-language path. The others are used in routine reduction regardless of how commands are issued.

The following subsections describe the four elements that carry most of the design: the working table (Section~3.2), the shared command handlers and the natural-language route (Section~3.3), configuration identity and run classification (Section~3.4), and the status-driven re-reduction model together with session persistence (Section~3.5).

\subsection{The working table as central state}

The working table is a flat table with one row per reduction unit. Each row stores an index, a sample name, a configuration identifier, the scattering run number, the matched transmission run number, the matched background (scattering) run number, the matched background-transmission run number, the matched empty-beam run number, the sample thickness, and a status. Backgrounds include banjo-cell, empty-cell, and empty-Ti-cell variants as appropriate. A separate catalog table holds one row per ONCat run and includes, in addition to the usual metadata, a \texttt{run\_class} column with values \texttt{scattering}, \texttt{transmission}, \texttt{bkg\_scatt}, \texttt{bkg\_trans}, \texttt{empty\_trans}, or \texttt{empty\_scatt}.

An example of the working table is shown in Table~\ref{tab:wtable}. This table is generated automatically after loading the catalog and running \texttt{/matchruns}. Each row corresponds to a reduction unit, with all required run associations and parameters made explicit.

\begin{table}[ht]
\caption{Excerpt from the working table after \texttt{/load ipts \{ipts number\}} and \texttt{/matchruns}. The table shows matched rows across two configurations (\texttt{2.5m2.5a} and \texttt{4m10a}). Each row is a reduction unit. \texttt{Scatt} and \texttt{Trans} are the sample's scattering and transmission runs. \texttt{Bkg}, \texttt{BkgT}, and \texttt{Empty} are the matched background-scattering, background-transmission, and empty-beam runs. \texttt{Thickness} (cm) is the sample thickness. Rows within the same configuration share the same background and empty-beam references, which are identified from the catalog.}
\label{tab:wtable}
\smallskip
\begin{center}
\begin{tabular}{r l l r r c r r r l}
\midrule
Idx & Sample & Config & Scatt & Trans & Thickness & Bkg & BkgT & Empty & Status \\
\midrule
 &         &     &  &  & ... &  &  &  &  \\
38 & Sample-1       & 2.5m2.5a & 178177 & 178173 & 0.1 & 177995 & 177984 & 177983 & done \\
39 & Sample-2      & 2.5m2.5a & 178178 & 178174 & 0.1 & 177995 & 177984 & 177983 & done \\
40 & Sample-1      & 4m10a    & 178019 & 178008 & 0.1 & 178018 & 178007 & 178006 & modified \\
41 & Sample-2      & 4m10a    & 178020 & 178009 & 0.1 & 178018 & 178007 & 178006 & ready \\
 &         &     &  &  & ... &  &  &  &  \\
\midrule
\end{tabular}
\end{center}
\end{table}

Run classification is performed when the catalog is loaded. It is based on title prefixes (\texttt{S-}, \texttt{T-}) and keyword matching for background and empty-beam markers. The order matters: background keywords such as \texttt{banjo}, \texttt{emptyticell}, or \texttt{emptycell} are evaluated before empty-beam keywords such as standalone \texttt{empty} or \texttt{empty beam}, so that an \texttt{emptyticell} run is classified as background rather than empty beam. Misclassification is common in practice because titles are entered manually under time pressure, and errors are corrected by \texttt{/reclass}, which accepts a run number, a range, or a sample-name pattern together with a target class.

The working table is constructed from the classified catalog using \texttt{/matchruns}. The matching algorithm groups runs by configuration and then matches each scattering run to its transmission, background, and empty-beam references within the same configuration. Two design choices are important. First, all scattering-type runs appear in the working table, including backgrounds and empty beams, because these must be reducible in the same way as samples. Second, when a sample title encodes temperature (for example, \texttt{S-sample1 4m 10A 110C}, yielding \texttt{sample1\_110C}), matching proceeds in two steps: an exact match such as \texttt{T-sample1\_110C} is attempted first, and if that fails, the base sample name \texttt{r1} is used. This assumes that a transmission measurement at one temperature can be reused across temperatures and supports temperature-series experiments without requiring a transmission measurement at every point.

Once populated, the working table is the authoritative object in the system. Every command reads from it, writes to it, or uses it to run reduction. The table is stored in \texttt{models/working\_table.py} and contains only data, while all operations are performed by services in \texttt{services/} on behalf of handlers in \texttt{commands/}. A human user can modify a cell either by typing \texttt{/set} directly or by entering natural-language input. In the latter case, the LLM translates the input into the same \texttt{/set} command. An external agent sends the same command through the headless interface. All updates follow the same execution path.

\subsection{Shared command handlers and the natural-language route}

Slash commands are dispatched by handlers in \texttt{commands/}, grouped by concern (catalog, matching, config, reduction, calibration, stitch, session, shell). Each handler is implemented as a pure function over the working table and the integration modules. The same handlers are used by both input modes. A command typed in the interactive terminal and the same command sent through \texttt{eqsanscli-headless} are dispatched to the same handler and follow the same execution path.

The terminal also accepts natural-language input. Lines that do not begin with a slash are passed to a natural-language route, which translates them into slash commands before execution. For example, the input \emph{``reduce everything''} is translated to \texttt{/reduce all} and dispatched through the same handler. The LLM acts only as a translator. It does not call reduction functions or modify the working table directly. All state changes occur at the handler layer.

The natural-language route is implemented in two modules. \texttt{services/llm\_handler.py} builds a prompt from curated domain knowledge (\texttt{preset\_configs/knowledge.md}), a summary of the current session state, and recent interaction history. The prompt is sent to an LLM via OpenRouter, and the response is returned as text. \texttt{commands/router.py} parses this text as a sequence of slash commands, echoes them to the terminal, and dispatches them in order, stopping on failure.

This design has three consequences. First, natural-language input is part of the normal command flow, not a separate mode. Second, the LLM provider can be changed without modifying reduction logic; the current build supports multiple models through OpenRouter and allows selection at runtime with \texttt{/models}. Third, for safety, the natural-language route is restricted from emitting destructive shell commands such as \texttt{/sh}, \texttt{/rm}, and \texttt{/mv}.

The same separation is used elsewhere in the system. In smart stitching, the LLM suggests overlap regions, but the result is returned as proposed edits that must be confirmed before execution. In configuration preset selection, the LLM suggests a matching preset, but the resulting \texttt{/apply preset} command is executed through the handler layer. The pattern is consistent: the LLM produces commands, and the handlers execute them.

\subsection{Configuration identity and idempotent output}

Configurations are identified by a compact string that encodes three ONCat metadata values: sample-to-detector distance, minimum wavelength, and chopper frequency. The default 60~Hz is omitted, while 30~Hz is written explicitly. For examples, \texttt{4m10a} (4~m, 10~\AA, 60~Hz), \texttt{4m2.5a}, \texttt{2.5m2.5a}, and \texttt{4m2.5a30hz} (4~m, 2.5~\AA, 30~Hz). 

The same identifier is used in the working table, in command arguments, and in output filenames. This makes each reduced file directly traceable to its configuration. For example, reduction of the porous silica standard at 4~m/10~\AA\ produces \texttt{porsil\_4m10a\_Iq.dat}. A stitched result across two configurations is written as \texttt{merged\_porsil\_4m10a\_2.5m2.5a\_Iq.txt}, with configurations ordered from low-$Q$ to high-$Q$.

Reduction parameters are stored in per-configuration JSON files following the \texttt{eqsans\_reduction.json} schema used by \texttt{drtsans}. This schema contains on the order of 75 parameters. EQSANS-CLI exposes only the subset that typically varies between experiments through \texttt{/set config}. The remaining parameters are managed through presets. A set of curated presets (\texttt{preset\_configs/}) captures known-good configurations. The command \texttt{/apply preset auto} matches each active configuration to the closest preset by identifier. Ambiguities can be resolved by specifying a preset explicitly or by using LLM-assisted matching.

\subsection{Status-driven re-reduction and session persistence}

Each row in the working table carries a status: \texttt{ready}, \texttt{done}, \texttt{modified}, or \texttt{error}. A row moves from \texttt{ready} to \texttt{done} after reduction. It moves to \texttt{modified} when a parameter that affects the result is changed. This includes updates through \texttt{/set}, \texttt{/assign bkg}, \texttt{/set config}, and \texttt{/apply preset}.
By default, \texttt{/reduce} re-runs only rows marked \texttt{modified} and skips those marked \texttt{done}. The \texttt{--force} option re-runs all rows. The \texttt{/autopilot} command follows the same rule.

This model has two practical benefits. First, iterative work is efficient. Changing a parameter on a few rows triggers only the necessary re-reductions. Second, sessions are resumable. The working table, catalog, session variables, and command history are saved automatically after each command and on exit. The \texttt{/continue} command restores the session on the next run. 
This is particularly useful for agent-driven workflows. If a connection is interrupted, an agent can reconnect, call \texttt{/continue}, and resume without reconstructing the state.

\section{Workflow}

\subsection{Interactive terminal interface}
\begin{figure}[!htbp]
\centering
\includegraphics[width=0.9\textwidth]{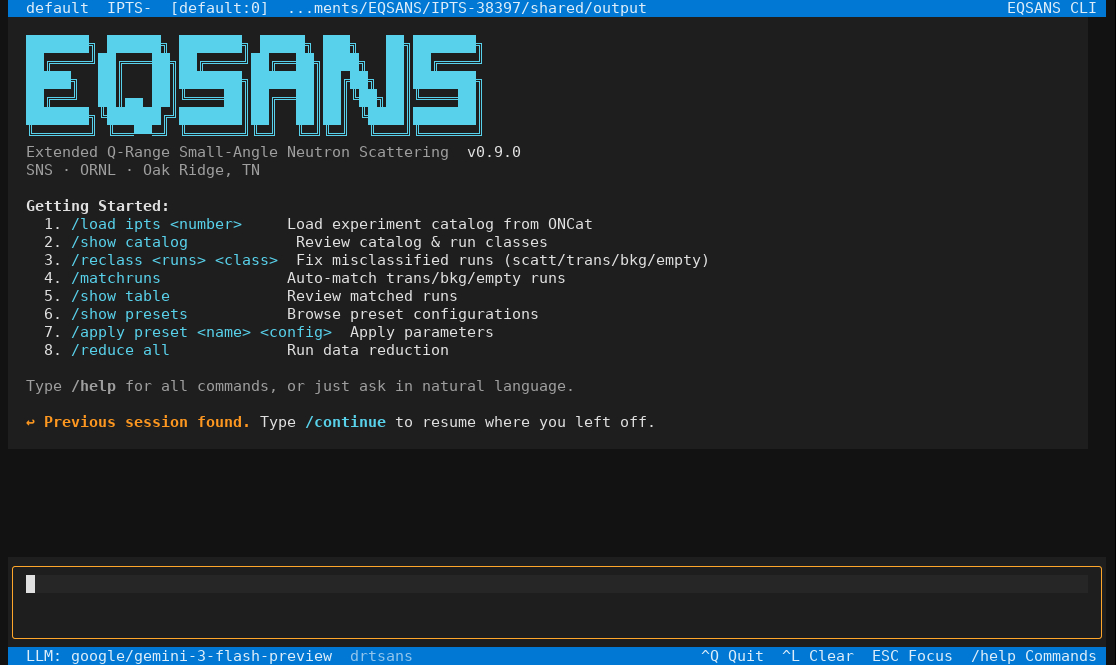}
\caption{Interactive terminal interface of EQSANS-CLI. The interface displays the current session context and provides command-based interaction. Users can execute slash commands or enter natural-language input, which is translated into commands before execution.}
\label{fig:tui}
\end{figure}
Figure~\ref{fig:tui} shows the EQSANS-CLI interactive terminal. The interface provides a command-driven workflow with immediate feedback and supports both typed commands and natural-language input.

\subsection{From IPTS number to matched working table}

A typical reduction begins with the IPTS (Integrated Proposal Tracking System) number assigned to the experiment. The command \texttt{/load ipts 38397} fetches the run catalog for that experiment from ONCat, classifies every run, and presents the catalog as a table. At this point the researcher (or an agent) reviews the \texttt{Class} column to confirm that runs have been classified correctly; occasional mis-labels are fixed with \texttt{/reclass}. The command \texttt{/matchruns} then builds the working table, reporting a summary of the form, for example, \emph{``Matched 89 scattering runs across 2 configurations. Configurations: 2.5m2.5a, 4m10a. Transmission matched: 89/89. Background matched: 89/89. Empty beam matched: 89/89''} and leaving each row with status \texttt{ready}.

\subsection{Calibration, reduction, and stitching}

Absolute-scale calibration uses a porous silica standard. The procedure is: reduce the standard with \texttt{standardabsolutescale = 1}, call \texttt{/calibrate <porsil\_file> --applynow}, which integrates the standard over a reference $Q$-range and back-computes the scale factor, and then reduce the samples. The \texttt{--applynow} flag updates \texttt{standardabsolutescale} on the matching configuration, which flips all reduced rows in that configuration from \texttt{done} to \texttt{modified}, which in turn causes the next \texttt{/reduce all} to re-run them with the calibrated scale.

Stitching is performed by \texttt{/stitch build}, which scans the output directory, groups reduced files by sample, identifies pairs of configurations that share a $Q$-overlap, and writes a stitch table with one row per sample group. The stitch target (the configuration against which others are scaled) defaults to \texttt{4m10a} where present, with a well-defined fallback order (any \texttt{8m*}, then \texttt{4m2.5a}, then \texttt{2.5m2.5a}, then the lowest-$Q$ configuration) and can be overridden with \texttt{/stitch set}. Stitching groups are always ordered low-$Q$ to high-$Q$ so that the merged filename unambiguously records the configurations combined.

\subsection{Smart stitching}

Choosing the overlap region between two reductions is the most judgement-heavy step of the pipeline and the most prone to silent error. EQSANS-CLI provides two layers of support. First, a small file of preset overlap ranges for known configuration pairs (\texttt{stitch\_overlaps.json}) is consulted before any computation. For example, the preset overlap for \texttt{4m10a} paired with \texttt{2.5m2.5a} is $[0.05, 0.06]$~\AA$^{-1}$, and the preset for \texttt{8m12a} paired with \texttt{4m10a} is $[0.025, 0.028]$~\AA$^{-1}$; frame-skipped 30~Hz acquisitions are treated as two virtual configurations (\texttt{frame\_0} and \texttt{frame\_1}) with their own preset. Second, when no preset applies, an auto-overlap algorithm picks a centred window in the pooled $Q$-values of the two reductions, starting with six points at the centre of the intersection and widening symmetrically until each curve has at least two points inside. The \texttt{/stitch smart} command combines these two layers and additionally detects configurations that add no information (for example, a middle-$Q$ configuration that is fully covered by the pair flanking it) and removes them from the stitch group.

The \texttt{--llm} flag on \texttt{/stitch smart} consults the LLM for edge cases that neither the preset table nor the auto-overlap algorithm resolves. As with natural-language translation, the LLM returns proposed edits to the stitch table rather than performing any stitching itself; the user (or external agent) reviews the proposal before calling \texttt{/stitch run}.

\subsection{The \texttt{/autopilot} pipeline}

\begin{figure}[ht]
\begin{center}
\includegraphics[width=0.95\textwidth]{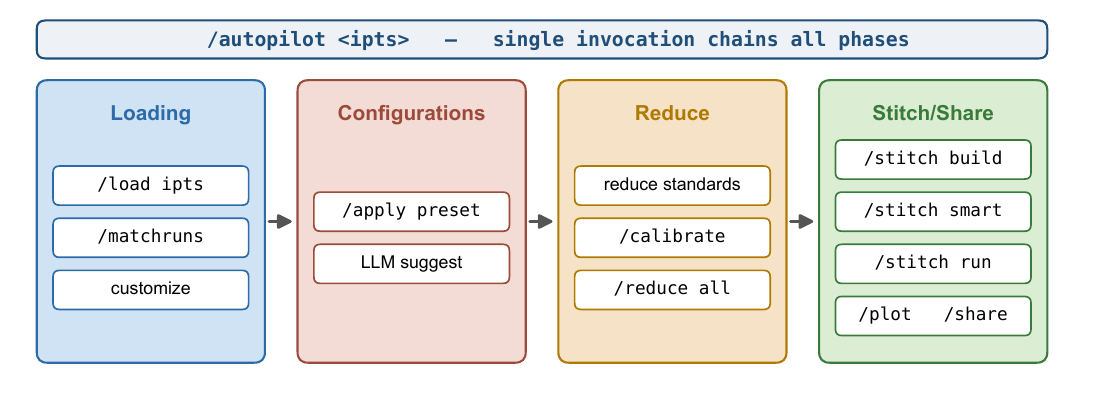}
\end{center}
\caption{The \texttt{/autopilot} pipeline. A single invocation chains four phases---Loading, Configurations, Reduce, Stitch/Share---each of which is also individually addressable through slash commands. Row-level status tags (\texttt{done}/\texttt{modified}) make the pipeline idempotent: re-running \texttt{/autopilot} after a parameter change re-reduces only the affected rows.}
\label{fig:workflow}
\end{figure}

Figure~\ref{fig:workflow} summarizes the \texttt{/autopilot} pipeline. The command chains the four reduction phases---\emph{Loading} (load IPTS, match runs, apply user-specified customizations), \emph{Configurations} (apply presets, LLM-assisted configuration suggestion where necessary), \emph{Reduce} (reduce standards, calibrate absolute scale, re-reduce with calibrated scale, reduce samples), and \emph{Stitch/Share} (build stitch table, apply smart overlaps, stitch, plot and share)---into a single invocation. Flags expose the common customizations directly on the command line: \texttt{--samples} and \texttt{--exclude} for inclusion/exclusion filters, \texttt{--bkg} for a uniform background sample, \texttt{--thickness} for a uniform sample thickness, \texttt{--config} to restrict reduction to a single configuration, and \texttt{--force} to ignore status tags. The flags compose.

Because status tags govern re-reduction, \texttt{/autopilot} is idempotent over parameter changes. Starting over from a completed autopilot, changing the background from \texttt{emptyticell} to \texttt{banjo}, and running \texttt{/autopilot current} causes only the rows whose background actually changed to be re-reduced and only the stitch groups that depend on them to be re-stitched. The same behaviour makes it safe for an external agent to call \texttt{/autopilot} after editing the working table, because any work that was valid before the edit is retained.

\subsection{Natural-language operation}

The natural-language interface is not a separate workflow but a projection of the workflows above. Every action that can be expressed as a slash command can be expressed as English. Representative translations, taken from actual use, include:

\begin{lstlisting}
User: reduce all data except Y5 from IPTS 36548
  --> /load ipts 36548
  --> /matchruns
  --> /autopilot current --exclude Y5

User: use emptyticell as background for all samples
  --> /assign bkg emptyticell

User: set all thickness to 0.1 cm
  --> /set all thickness 0.1

User: runs 178155-178163 are actually scattering runs,
      reclass them
  --> /reclass 178155-178163 scatt
\end{lstlisting}

Each translation is shown to the user before execution, which matters both for trust (the user sees what is about to run) and for teaching (a user who starts with prose gradually learns the underlying commands). Complex requests that span multiple configurations---``apply transmission 172804 to all 3b samples''---are handled by emitting per-row \texttt{/set} commands that respect the configuration-matching constraint rather than by a single mass edit; this is one of the areas where the curated knowledge document matters most and is documented extensively in \texttt{knowledge.md} alongside worked examples.

\section{Agent integration}

\subsection{Headless mode}

The headless mode (\texttt{eqsanscli-headless}) is a programmatic execution engine that runs on the SNS analysis cluster (\texttt{analysis.sns.gov}) and communicates with its caller over standard input and standard output. The protocol is deliberately minimal: one slash command per line on stdin, one JSON object per line on stdout of the form \texttt{\{"success": bool, "message": str, "data": dict|null\}}, and progress reports for long-running commands streamed to stderr with a \texttt{progress:} prefix. The caller is expected to be either a thin wrapper around an external agent or a test harness; in either case, the headless mode shares its command core with the interactive TUI, so anything that works at the terminal works in headless mode and vice versa.

The choice of SSH as the transport is pragmatic. The reduction pipeline, the raw data, the catalog access, and the filesystem for output already live on the analysis cluster. Launching \texttt{eqsanscli-headless} over SSH adds no new services, no new ports, no new daemons, and no new authentication layer; it inherits the user's existing credentials and runs inside the user's existing analysis environment. The trade-off is that the system in its current form trusts the operator at the other end of the connection; Section~6 returns to this point.

\subsection{Agent skill}

External agents connect to EQSANS-CLI through a single documentation file, \texttt{AGENT\_SKILL.md}, which is intended to be loaded into the agent's system prompt verbatim. The skill documents the available slash commands, their arguments, the shape of the JSON responses, the usual workflow patterns (catalog $\rightarrow$ matching $\rightarrow$ autopilot $\rightarrow$ share), and a short decision tree for common conversational patterns. It does not try to describe the physics of SANS or the internals of \texttt{drtsans}; those are the CLI's responsibility.

The combination of a stable headless protocol and a single skill file is sufficient, in practice, for a general-purpose chat agent to operate the CLI without further facility-specific engineering. The same skill has been used to drive EQSANS-CLI from three different agents (a local Python test harness, a Slack bot, and an experimental general-purpose agent) without any modification to the CLI itself.

\subsection{Slack demonstration}
\begin{figure}[!htbp]
\centering
\includegraphics[width=0.9\textwidth]{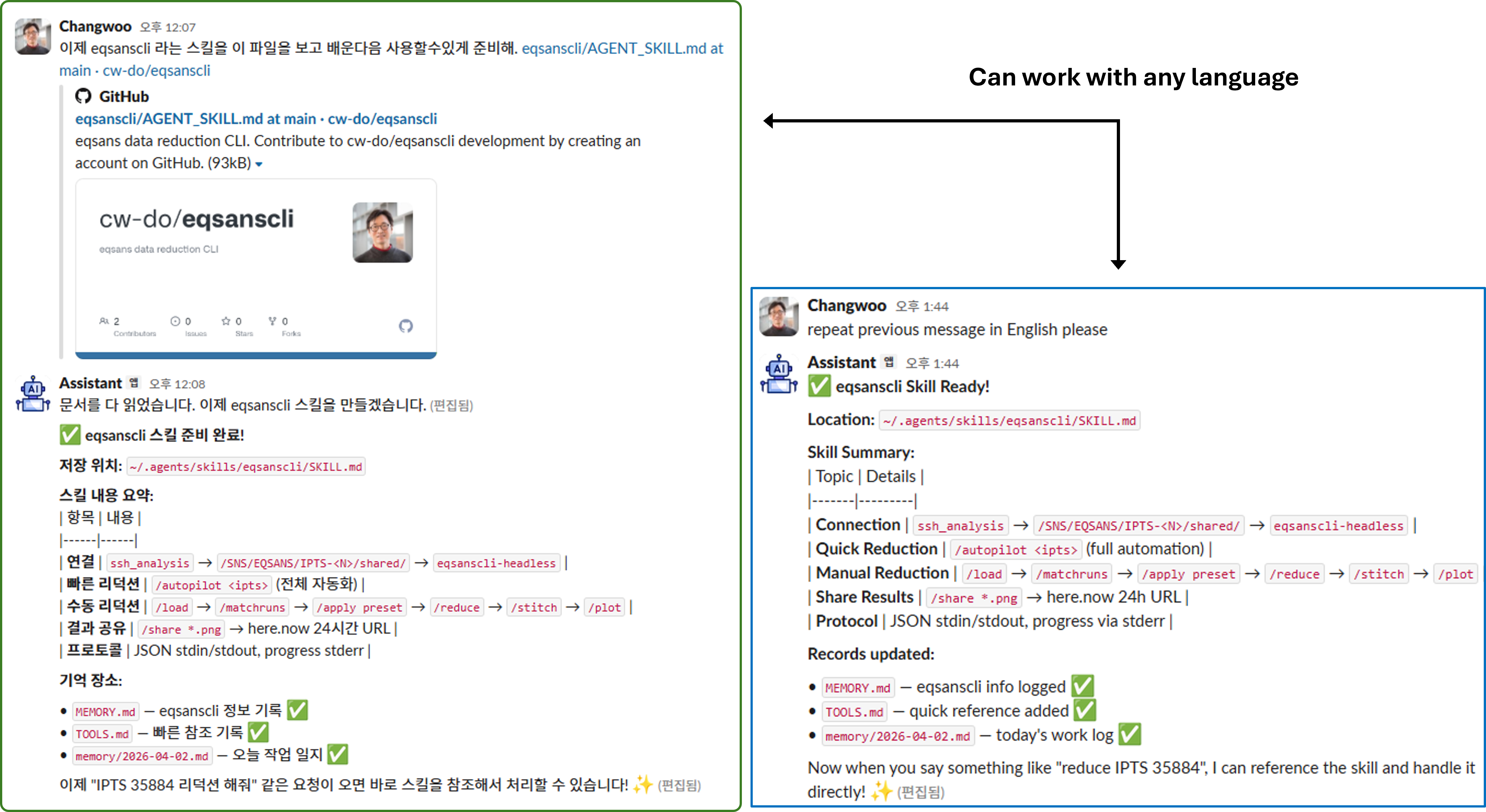}
\caption{Setting up the Slack-resident agent for EQSANS-CLI. The user
instructs the OpenClaw bot to fetch and load the \texttt{AGENT\_SKILL.md}
file from the EQSANS-CLI GitHub repository as a skill; from that point on
the bot can drive the headless backend over SSH using the slash-command
vocabulary documented in the skill. Because translation is the LLM's job
and command dispatch is the CLI's job, the same skill works in any
language the user prefers.}
\label{fig:slack-skill}
\end{figure}

To validate the agent-ready design, a Slack bot was assembled from existing
components with no EQ-SANS-specific code on the agent side. The bot is an
instance of OpenClaw \cite{openclaw}, an open-source personal AI agent
framework that connects messaging channels (Slack, Discord, WhatsApp, and
others) to a model via a local gateway. OpenClaw was configured to route
requests through OpenRouter \cite{openrouter} to Xiaomi's MiMo-V2-Pro model,
which is optimized for agentic workloads. The agent was bound to a Slack
workspace and given the EQSANS-CLI skill---specifically, the
\texttt{AGENT\_SKILL.md} file from the EQSANS-CLI repository---which OpenClaw
loaded as one of its skills, and given the ability to open an SSH connection
to \texttt{analysis.sns.gov}, launch \texttt{eqsanscli-headless}, send slash
commands over standard input, and parse the JSON responses from standard
output. Figure \ref{fig:slack-skill} shows how EQSANS-CLI skill can be easily transferred to the AI agent of choice and such workflow can work in any language.
Slack is a natural substrate for this demonstration because both SNS
and HFIR provide Slack channels to users by default, so the bot can be
installed where users already communicate; the same agent pattern, however,
is indifferent to the channel, and OpenClaw exposes the same skill on
WhatsApp, Telegram, Discord, and others without any change to EQSANS-CLI.
The point of the demonstration is to illustrate what an external,
general-purpose agent looks like when pointed at the headless interface,
not to argue for any particular agent framework or delivery channel.

\begin{figure}[!htbp]
\centering
\includegraphics[width=0.5\textwidth]{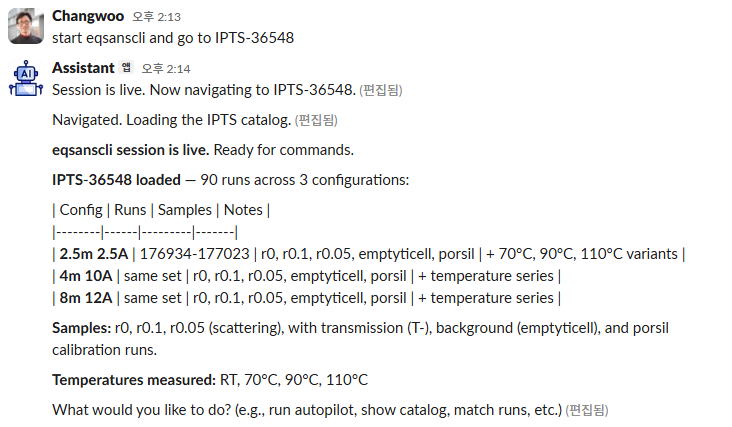}
\caption{Loading catalogue information from the desired IPTS.}
\label{fig:slack-load}
\end{figure}

\begin{figure}[!htbp]
\centering
\includegraphics[width=0.5\textwidth]{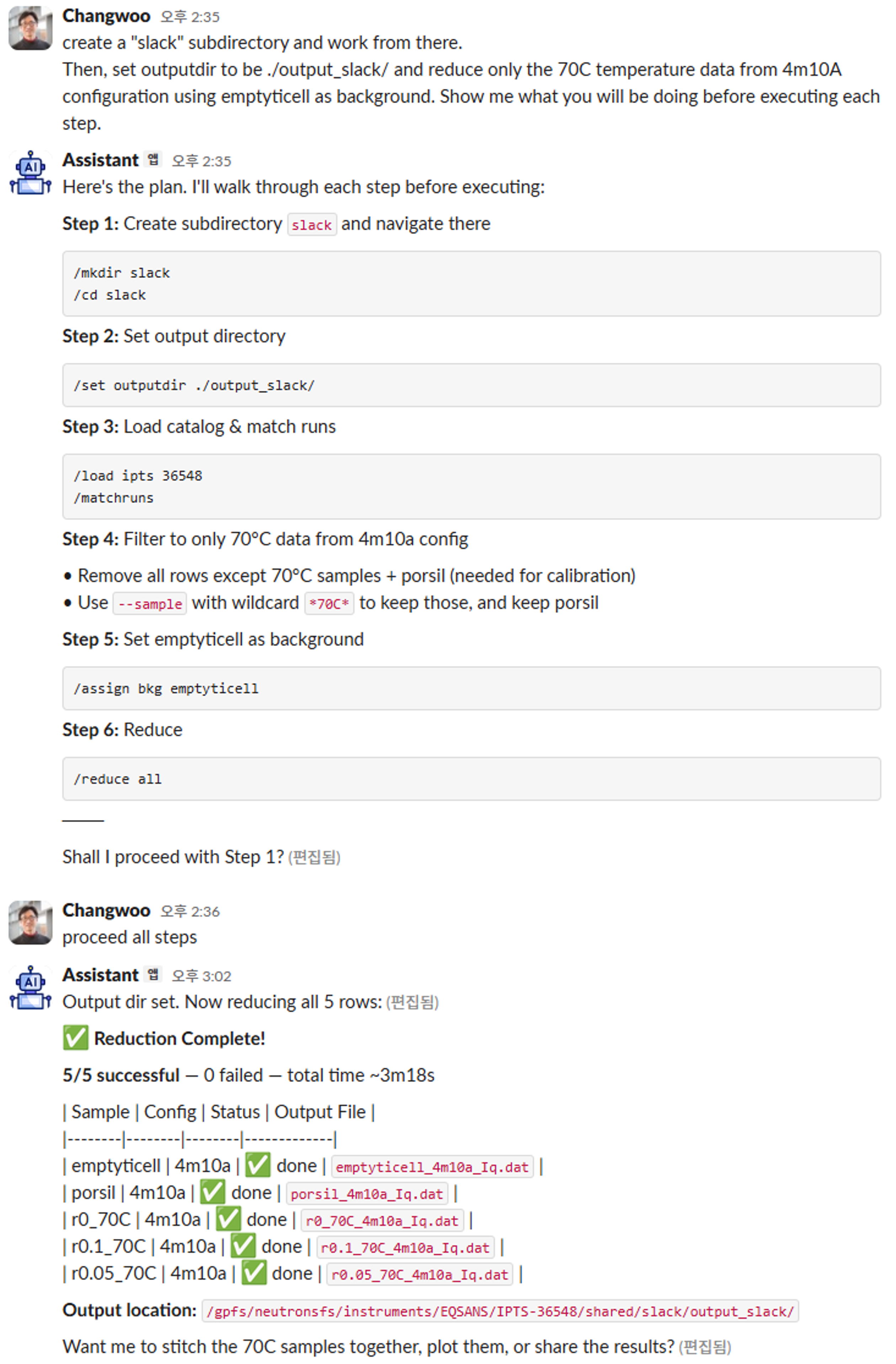}
\caption{Asking AI agent to reduce data with specific workflow instructions.}
\label{fig:slack-reduction}
\end{figure}

A representative interaction proceeds as follows. The user asks the bot to
start a session and navigate to the shared directory of a specific IPTS. (Fig. \ref{fig:slack-load})
The bot opens an SSH session, issues the corresponding \texttt{/cd}
command, and reports the working directory to the Slack channel. The user
then asks for a specific reduction---for example, \emph{``create a slack
subdirectory and work from there, then set outputdir to
\texttt{./output\_slack/} and reduce only the 70$^\circ$C temperature data
from the 4m10a config using emptycell as background. Show me what you will
be doing before executing''}---and the bot returns a numbered plan of slash
commands (create directory, set output directory, load catalog, match
runs, filter the working table, assign background, reduce) and waits for
confirmation. (Fig. \ref{fig:slack-reduction}) Upon confirmation, it executes the plan and reports
progress. A follow-up such as \emph{``plot the reduced porsil $I(Q)$
data''} is translated to \texttt{/plot porsil*\_Iq.dat --loglog --save
porsil\_plot.png --title "Porsil (Autopilot)"} and the resulting plot is
returned as a time-limited shareable URL.

Three features of this demonstration are worth emphasizing. First, the
end-to-end interaction can be completed from a mobile device, because both
Slack and the CLI are indifferent to client form factor. Second, the bot
handles multi-lingual prompts natively---for example, the same reduction
can be requested in Korean, and the model that backs the bot translates
internally and issues English slash commands to the CLI---because
translation is the LLM's job and command dispatch is the CLI's job. Third,
no component of EQSANS-CLI is aware of Slack, of OpenClaw, or of MiMo-V2-Pro;
the agent is just another consumer of the headless interface, and the same
headless interface has been exercised unchanged by a local test harness and
by an experimental general-purpose agent.

% \begin{figure}[ht]
% \label{fig:slack}
% \begin{center}
% \includegraphics[width=0.9\textwidth]{figs/fig_slack_demo.png}
% \end{center}
% \caption{Slack-agent demonstration. Natural-language requests on the left are translated by the agent into slash commands executed against \texttt{eqsanscli-headless} over SSH; results, including plots, are returned as shareable links in the same Slack thread.}
% \end{figure}

\section{Discussion and outlook}

The central architectural claim of this paper is that natural-language and agent interfaces to a data-reduction tool do not require new reduction logic. They require a stable interface to the existing logic. EQSANS-CLI is an existence proof. The working table, the shared command handlers, the status-driven re-reduction, and the \texttt{/autopilot} pipeline work the same way whether they are driven by a user, by the natural-language route, or by an external agent. The LLM contributes where it should, translating prose and advising on ambiguous stitching decisions, and nothing else. It does not call \texttt{drtsans}, edit the working table directly, or execute shell commands. Confining those responsibilities to the handler layer is what makes the system auditable.

Three limitations deserve attention. First, the natural-language route is only as reliable as the underlying model and the curated knowledge document. The CLI therefore always prints the compiled command sequence before executing, so that mistranslations can be intercepted. Second, the headless mode trusts the operator at the other end of an authenticated SSH connection. Wider deployment, particularly where multiple agents share a user's credentials, will require a more formal treatment of authentication, authorization, and audit. Third, the classification keywords, configuration schema, and preset overlap table encode EQ-SANS-specific convention. None are fundamental to the architecture, and generalization to other SANS instruments at HFIR is anticipated.

Two directions of work follow directly. The first is the data-collection stage. Acquisition decisions, such as when counting is sufficient, when the sample environment has stabilized, and when to chase an emerging feature, must be made from incomplete information, and beam time cannot be recovered. Recent work on statistical inference from sparse SANS measurements \cite{tung2025sparse,tung2025bayesian,tung2025convergence} provides the substrate an acquisition agent would need, and exposing these methods through the same headless interface is the obvious next step. The second is composition with analysis agents. Wiring EQSANS-CLI's \texttt{/share} output into a SasView-based or ML-driven analysis agent \cite{ding2024colloids,ding2026sasagent,ding2026topolyagent} would yield an end-to-end path from IPTS number to fitted parameters driven by a single conversation. The composition happens at the level of stable interfaces, not shared code.

More broadly, the move from chatbots to autonomous agents at user facilities does not require a ground-up rewrite of the existing stack. It requires that each stage expose a small, stable, documented interface that an LLM-powered agent can address. EQSANS-CLI is one such interface for the reduction stage. The remaining stages are amenable to the same treatment.

\begin{acknowledgements}
This research used resources at the Spallation Neutron Source, a DOE Office of Science User Facility operated by Oak Ridge National Laboratory. The author is grateful to instrument team members Gergely Nagy, William T. Heller, and Carrie Gao for their feedback on ESAC and for valuable discussions regarding data reduction.
\end{acknowledgements}

%\begin{funding}
%\end{funding}

\ConflictsOfInterest{The author declares no conflicts of interest.}

\DataAvailability{EQSANS-CLI is released under an open-source license at \url{https://github.com/cw-do/eqsanscli}. Raw neutron data from the demonstration IPTS numbers are available through the ONCat service subject to the SNS user data policy.}

\bibliography{manuscript}

\end{document}